# S3PET: Semi-supervised Standard-dose PET Image Reconstruction via Dose-aware Token Swap

Jiaqi Cui
School of Computer Science
Sichuan University
Chengdu, China
jiaqicui2001@gmail.com

Pinxian Zeng
School of Computer Science
Sichuan University
Chengdu, China
651215874@qq.com

Yuanyuan Xu
School of Computer Science
Sichuan University
Chengdu, China
YuanyuanXuSCU@outlook.com

Xi Wu
School of Computer Science
Chengdu University of Information Technology
Chengdu, China
xi.wu@cuit.edu.cn

Jiliu Zhou
School of Computer Science
Sichuan University
Chengdu, China
zhoujiliu@cuit.edu.cn

Yan Wang*
School of Computer Science
Sichuan University
Chengdu, China
wangyanscu@hotmail.com

*Abstract*—To acquire high-quality positron emission tomography (PET) images while reducing the radiation tracer dose, numerous efforts have been devoted to reconstructing standard-dose PET (SPET) images from low-dose PET (LPET). However, the success of current fully-supervised approaches relies on abundant paired LPET and SPET images, which are often unavailable in clinic. Moreover, these methods often mix the dose-invariant content with dose level-related dose-specific details during reconstruction, resulting in distorted images. To alleviate these problems, in this paper, we propose a two-stage Semi-Supervised SPET reconstruction framework, namely S3PET, to accommodate the training of abundant unpaired and limited paired SPET and LPET images. Our S3PET involves an un-supervised pre-training stage (Stage I) to extract representations from unpaired images, and a supervised dose-aware reconstruction stage (Stage II) to achieve LPET-to-SPET reconstruction by transferring the dose-specific knowledge between paired images. Specifically, in stage I, two independent dose-specific masked autoencoders (DsMAEs) are adopted to comprehensively understand the unpaired SPET and LPET images. Then, in Stage II, the pre-trained DsMAEs are further finetuned using paired images. To prevent distortions in both content and details, we introduce two elaborate modules, i.e., a dose knowledge decouple module to disentangle the respective dose-specific and dose-invariant knowledge of LPET and SPET, and a dose-specific knowledge learning module to transfer the dose-specific information from SPET to LPET, thereby achieving high-quality SPET reconstruction from LPET images. Experiments on two datasets demonstrate that our S3PET achieves state-of-the-art performance quantitatively and qualitatively.

## I. INTRODUCTION

As an advanced nuclear imaging technique, positron emission tomography (PET) enables the visualization of the metabolic process of the human body, and has been widely applied in hospitals for diagnosis and treatment [1,2]. In clinic, high-quality PET images are acquired by injecting a sufficient (standard) dose of radioactive tracers into the human body, which inevitably poses potential health risk. Lowering the injected tracer dose, on the other hand, results in degraded image quality and limited diagnostic information [3]. As shown in Fig. 1, standard-dose PET (SPET) images contain more details and less noise, whereas low-dose PET (LPET) images appear blurry and noisy. Therefore, to reduce the radiation hazards while ensuring the image quality, it is of great interest to reconstruct SPET images from the corresponding LPET images.

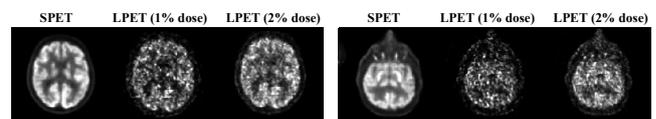

Fig. 1. Two typical cases of paired SPET images and the corresponding 1% dose, 2% dose LPET images. Compared to SPET images, LPET images contains more noises and less details.

The advent of deep learning has spurred a plethora of endeavors towards deep learning-based PET reconstruction [4-19]. Specifically, inspired by the remarkable performance of the generative adversarial network (GAN) in image synthesis, Wang *et al.* [5] introduced the adversarial training strategy to PET reconstruction and accordingly presented a 3D conditional GAN. More recently, taking the advantage of transformer in capturing long-range interactions, Luo *et al.* [10] embedded the trans-former into a convolutional neural network (CNN)-based framework to preserve both global semantic dependencies and low-level local connectivity in PET reconstruction. Despite their commendable performance, these approaches heavily rely on the avail-ability of paired SPET and LPET images, which are challenging to collect in routine clinical practice.

To alleviate the demand for paired low- and standard-dose images in the field of medical image reconstruction, some efforts have turned to semi-supervised learning which involves utilizing both the rare paired and the abundant unpaired low- and standard-dose images for reconstructing high-quality standard-dose images [20, 21]. For example, Wang *et al.* [21] designed a semi-supervised network for computed tomography (CT) metal artifact reduction, where the clean CT images are learned from the generator. Along the research direction of PET image reconstruction, the semi-supervised scenario remains largely unexplored. The pioneering effort [22] developed a semi-supervised PET reconstruction framework that employs a structural

This work is supported by the National Natural Science Foundation of China (NSFC 62371325, 62071314), Sichuan Science and Technology Program 2023YFG0263, 2023YFG0025, 2023NSFSC0497.
The first two authors contribute equally to this work.
*Corresponding author.

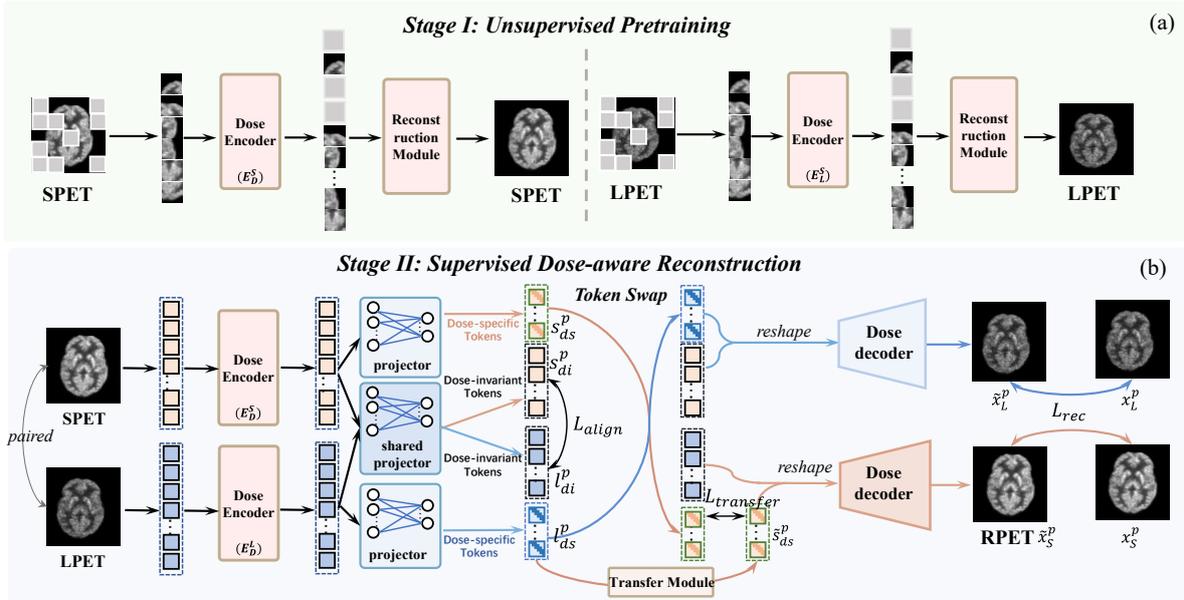

Fig. 2. Illustration of the proposed S3PET. It comprises (a) an unsupervised an unsupervised pre-training stage (Stage I) aimed at comprehensively capturing the dose knowledge from unpaired LPET and SPET images, respectively, and (b) a supervised dose-aware reconstruction stage (Stage II) focused on effectively transferring the dose-specific knowledge from SPET to LPET for generating RPET images.

consistency loss to maintain the coherence of structure details in the mapping from LPET to SPET.

However, by directly translating low-dose images to standard-dose ones, these methods fail to fully explore the dose-invariant content shared among LPET and SPET, as well as the dose-specific details related to particular tracer dose level (i.e., SPET contains less noise, whereas LPET exhibits contrasting characteristics) between LPET and SPET images. Such neglect can impede the learning of the intricate relationships between SPET and LPET images under limited paired images, distorting both content and details during reconstruction. Moreover, given the shared dose-invariant content and distinctly possessed dose-specific details between LPET and SPET images [23], we contend that transferring dose-specific knowledge from SPET to LPET is sufficient for achieving the LPET-to-SPET reconstruction. Therefore, a primary concern lies in how to reserve the dose-invariant information while acknowledging and transferring the dose-specific knowledge between SPET and LPET images for enhancing the performance of semi-supervised PET reconstruction.

Motivated to address the above challenges, in this paper, we propose a two-stage Semi-Supervised framework for high-quality SPET image reconstruction, namely S3PET. Our S3PET model contains an unsupervised pre-training stage (Stage I) and a supervised dose-aware reconstruction stage (Stage II) to accommodate the training with abundant unpaired and rare paired PET images, respectively. Specifically, in Stage I, we are inspired by masked image modeling [24] and adopt two independent dose-specific masked autoencoder (DsMAEs) respectively for unpaired LPET and SPET images to comprehensively capture their corresponding dose knowledge. The well pre-trained DsMAEs can serve as strong feature extractors for subsequent reconstruction. Then, in Stage II, we finetune the two pre-trained DsMAEs on the paired images. Particularly, to facilitate an effectively LPET-to-SPET reconstruction while preventing distortion in content and details, we design a dose knowledge decouple (DKD) module to disentangle the respectively dose-specific and dose-invariant knowledge of LPET and SPET, and a dose-specific knowledge learning (DKL) module to transfer the dose-specific information from SPET to LPET. The main contributions of this paper are summarized as follows:

(1) We propose S3PET, a novel two-stage semi-supervised learning framework that utilizes sufficient unpaired and rare paired SPET and LPET images for high-quality SPET reconstruction.

(2) We introduce an unsupervised pre-training stage that leverages two independent dose-specific masked autoencoders (DsMAEs) to comprehensively undermine the varied dose knowledge in unpaired LPET and SPET images, thereby providing strong feature extractors for the following reconstruction.

(3) To prevent distortion of both content and details, we design a dose-aware reconstruction stage that incorporates a dose knowledge decouple (DKD) module and a dose-specific knowledge learning (DKL) module to respectively disentangle and transfer the dose-specific information between paired SPET and LPET images, thus achieving high-quality SPET reconstruction.

(4) Extensive experiments on two public PET datasets demonstrate that our approach outperforms the state-of-the-art PET reconstruction approaches in both qualitative and quantitative measures.

## II. RELATED WORKS

### A. Deep Learning in PET Reconstruction

In the past decade, numerous deep learning-based methods have been proposed to reconstruct clinically approved SPET images from the corresponding LPET images. Along this research direction, numerous works adopt the CNN to realize the transformation from LPET to SPET images. For example, Spuhler *et al.* [8] designed a dilated convolutional neural network which allows the network to perceive larger and more robust feature representations, thereby achieving superior performance. Then, appealing to the power of GAN in image generation, efforts have been dedicated to leveraging GAN-based networks for SPET reconstruction. Particularly, Wang et al. [5] injected the spirit of adversarial training into 3D U-Net and designed a 3D conditional GAN for high-quality PET estimation. Zhou et al. [13] adopted a task-driven strategy and accordingly devised a segmentation-guided style-based GAN to predict SPET images at low dose. Recently, some studies have applied transformer for SPET estimation, due to their advantage in modeling long-range global information. For instance, Luo et al. [10] embedded transformer into the U-Net bottleneck to capture both the global semantic dependencies and the low-level spatial information. Following this, Cui et al. [14] developed a cascaded transformer for SPET reconstruction. However, these methods typically rely on paired PET images, which are hard to collect and are rare in real clinical scenarios. Facing this problem, we follow the philosophy of semi-supervised learning and propose a novel framework to accommodate both sufficient unpaired and rare paired SPET and LPET images for SPET reconstruction.

*B. Masked Image Modeling in Medical Images*

Masked Image Modeling (MIM) is a potent self-supervised learning paradigm that has garnered increasing interest recently. By reconstructing the masked portion of images, models can learn informative feature representations beneficial for various downstream tasks. In the field of medical images, various works [24-27] have explored the effectiveness of MIM pretraining. For example, Zhou et al. [25] applied the masked autoencoder (MAE) pre-training paradigm to improve diverse medical image tasks such as chest disease classification, abdominal multi-organ segmentation, and brain tumor segmentation. Wang et al. [27] redesign the classical MAE based on SwinIR to enable the exploitation of unlabeled data and facilitate structural preservation for the low-dose computed tomography (CT) image denoising. Appealed by the excellent feature representation ability of MIM, we design the dose-specific masked autoencoders (DsMAEs) to comprehensively undermine the varied dose knowledge in unpaired LPET and SPET images, thereby providing strong feature extractors for the following reconstruction.

## III. METHODOLOGY

The architecture of the proposed S3PET is depicted in Fig. 2, which consists of an unsupervised pre-training stage (Stage I) and a supervised dose-aware reconstruction stage (Stage II). Overall, the unpaired PET images are first utilized to train their independent dose-specific masked autoencoders (DsMAEs) in Stage I. Then, the limited paired PET images are employed in Stage II to fine-tune the pre-trained DsMAEs and facilitate an effective dose-specific knowledge transfer between SPET and LPET, thereby obtaining the reconstructed PET (RPET) images. Details of each stage are given in the following sub-sections.

*A. Unsupervised Pre-training Stage*

As shown in Fig. 2(a), we pre-train two independent DsMAEs for unpaired LPET and SPET images, respectively. Following [24], each DsMAE consists of three sub-parts, i.e., a masking module, a dose encoder, and a reconstruction module. Specifically, the masking module first randomly masks a portion of the input images. For description, we denote an input unpaired PET image as $x_D^u \in \mathbb{R}^{C \times H \times W}$, where $H$, $W$, and $C$ represent the height, weight, and channel dimension, and $D = \{L, D\}$ corresponds to LPET or SPET. Given $x_D^u$, we divide the image into $N$ non-overlapping patches, and then randomly sample $r\%$ patches while masking the remaining ones. Note that, considering that SPET images are larger in quantity than LPET images in clinic [22], we mask LPET images at a higher rate than that of SPET images, thus forcing the network to learn more representative features with fewer images. Then, a transformer-based dose encoder is applied to capture the image representation, enabling the network to have a holistic understanding of the dose knowledge. Taking the masked image patches as input, the dose encoder, denoted as $E_D^u$, embeds these patches via a linear projection with added positional embeddings, and then processes the resulting features via $T$ transformer blocks. After that, the reconstruction module recovers the masked patches to realize the reconstruction. A light-weight linear layer is adopted as the decoder to achieve the prediction.

By masking out a portion of the image and predicting the missing components, the complex functional and structural interactions in the image can be fully excavated [19], thereby allowing the dose encoders $E_D^u$ to effectively capture the varied dose knowledge in unpaired LPET and SPET images, respectively.

*B. Supervised Dose-aware Reconstruction Stage*

The supervised dose-aware reconstruction stage is designed to transfer the dose-specific knowledge from SPET to LPET in paired images, thereby achieving SPET reconstruction. As depicted in Fig. 2(b), this stage adopts a bilateral network including a main branch that reconstructs SPET images from LPET and an auxiliary branch to transfers the dose-specific knowledge from SPET to LPET. Particularly, the two branches comprise two independent encoders respectively initialized with the pre-trained dose encoder $E_D^L$ (for LPET) and $E_D^S$ (for SPET). A dose knowledge decouple (DKD) module and a dose-specific knowledge learning (DKL) module are shared between two branches. In addition, a transfer network is embedded in the master branch to learn the mapping of dose-specific knowledge from LPET images to SPET images. Finally, two independent linear-based dose decoders are injected in the auxiliary branch and main branch for LPET synthesis and SPET reconstruction, respectively.

**Dose Knowledge Decouple (DKD).** Given the paired LPET and SPET images $x^p = \{x_L^p, x_S^p\} \in \mathbb{R}^{2 \times C \times H \times W}$, they are first fed into the corresponding dose encoders $E_D^L$ and $E_D^S$ to obtain token embeddings, denoted as $e^p = \{e_L^p, e_S^p\} \in \mathbb{R}^{2 \times N \times d}$, where $N$ is the number of tokens and $d$ is the dimension of the embedding. Then, the DKD module takes $e^p$ as input, and produce the dose-specific and dose-invariant tokens of LPET (i.e., $\{l_{ds}^p, l_{di}^p\}$) and SPET (i.e., $\{s_{ds}^p, s_{di}^p\}$) respectively via linear projector, where $l_{ds}^p$ and $s_{ds}^p$ are dose-specific ones, and $l_{di}^p$ and $s_{di}^p$ are dose-invariant ones. To encourage the excavation of more dose-invariant tokens, the weights of dose-invariant projectors are shared. Meanwhile, a dose-invariant alignment constraint is imposed to maximizes the shared invariant information. By doing so, the dose-specific knowledge can be fully decoupled from the dose-invariant ones.

**Dose-specific Knowledge Learning (DKL).** We contend that if the dose-specific and dose-invariant information can be fully decoupled, swapping the respective dose-specific tokens of LPET and SPET can realize the translation between LPET and SPET images. Therefore, we first swap the dose-specific tokens between SPET and LPET via a token swap operation. These swapped dose-specific tokens, along with their dose-invariant counterparts are sent to separate dose decoders to generate synthesis LPET images and RPET images, respectively. To effectively learn such mapping between LPET and SPET from the perspective of dose-specific tokens and facilitate an effective dose knowledge transfer, we design a linear-based transfer module aimed at learning the translation from $l_{ds}^p$ to $s_{ds}^p$, during the transfer of dose-specific knowledge between SPET and LPET images.

**Dose Decoders.** During model training, we find that bluntly and brutally stitching the output sequential tokens together according to the spatial location will result in blurred and unnatural reconstructions. We argue that the reason behind this is that the pure transformer-based encoders lack spatial local information required in dense pixel-level regression tasks such as PET reconstruction. Therefore, we employ two CNN-based dose decoders to capture the spatial local information during reconstruction. The two dose decoders share the same structures consisting of four convolution blocks, each structured as Convolution-BatchNorm-ReLU. Three Up-sampling layers are embedded among blocks. All the tokens are first reshaped to 2D feature maps and then delivered to the decoders. For the auxiliary branch, it receives the feature map from the dose-invariant token $s_{di}^p$ and dose-specific token $l_{ds}^p$ to synthesis the predicted LPET images (denoted as $\tilde{x}_L^p$), while the master branch takes both the transferred dose-specific tokens $\tilde{s}_{ds}^p$, along with the dose-invariant tokens $l_{di}^p$ to predict the final RPET images (denoted as $\tilde{x}_S^p$).

### C. Objective Function

**Unsupervised Pre-training Stage.** In this stage, an image reconstruction loss is employed for supervision. It aims to narrow the gap between the reconstructed result and the input. Specifically, the former loss applies the L1 loss as the image reconstruction loss to encourage less blurring, which can be expressed as follows:

$$L_{StageI} = E_{\tilde{x}_D^u, x_D^u}[||\tilde{x}_D^u - x_D^u||_1], \quad (1)$$

**Supervised Dose-aware Reconstruction Stage.** For this stage, the objective function consists of three parts, i.e., a dose-invariant alignment loss, a transfer loss, and a reconstruction loss. Specifically, the dose-invariant alignment loss employs the $JS$-divergence $JS(\cdot)$ to minimize the distribution gap between $l_{di}^p$ and $s_{di}^p$, defined as follows:

$$L_{align} = JS(l_{di}^p, s_{di}^p), \quad (2)$$

To ensure an effective transfer of dose-specific knowledge, we also leverage the $JS(\cdot)$ as the transfer loss between the swapped dose-specific token $s_{ds}^p$ and the predicted dose-specific token $\tilde{s}_{ds}^p$, which can be expressed as:

$$L_{transfer} = JS(s_{ds}^p, \tilde{s}_{ds}^p), \quad (3)$$

For the reconstruction loss, we also apply the L1 loss as to narrow the gap between the reconstructed image and the input image in both the auxiliary and the master branches, thereby enhancing the dose knowledge transfer while enhancing the reconstruction quality. It can be described as the following $\gamma$-weighted sum:

$$L_{rec} = E_{\tilde{x}_L^p, x_L^p}[||\tilde{x}_L^p - x_L^p||_1] \quad (4)$$
$$+ \gamma E_{\tilde{x}_S^p, \tilde{x}_S^p}[||\tilde{x}_S^p - x_S^p||_1],$$

where $x_L^p$ and $x_S^p$ represent the real LPET and SPET images.

Overall, the final objective function of Stage II is formulated as follows:

$$L_{StageII} = L_{align} + \lambda_1 L_{transfer} + \lambda_2 L_{rec}. \quad (5)$$

where $\lambda_1$ and $\lambda_2$, are the weighting coefficient to balance these three terms.

## IV. EXPERIMENTAL AND RESULTS

### A. Experimental Settings

**Datasets.** We train and validate our model on a publicly available Ultra-low Dose PET Imaging Challenge (UDPET) dataset [28] In this dataset, all 18F-FDG PET imaging data is acquired using the Siemens Biograph Vision Quadra system. The LPET is produced by subsampling a portion of the SPET scan. Particularly, LPET images with the dose reduction factor (DRF) of r0 (i.e., 1/r0 doses to SPET images) are employed, aiming to validate the feasibility and effectiveness of our method under the challenging ultra-low-dose scenario. We select a total number of 120 unpaired SPET images and 50 paired LPET and SPET images for experimentation. In Stage I, 30 LPET images (from 50 paired LPET and SPET images) and 120 unpaired SPET images are used to train their respective DsMAE networks, thus obtaining the pre-trained dose encoders $E_L^S$ and $E_D^S$. In Stage II, the 30 selected LPET images and the corresponding paired SPET images are used to further finetune the dose encoders $E_L^S$ and $E_D^S$, as well as the DKD module and the DKL module, thus generating the RPET images. The rest 20 paired LPET and SPET images are used for evaluation. Moreover, to prevent over-fitting problem and increase the training samples, each 3D scan of size 128 × 128 × 128 are split into 128 2D slices of size 128 × 128.

**Implementation Details.** The proposed method is implemented by the Pytorch framework using an NVIDIA GeForce RTX 3090 GPU with 24GB memory. In Stage I, we train the knowledge encoders for 300 epochs with a batch size of 32 and a learning rate of 2e-4. In Stage II, the network is trained for 100 epochs with a batch size of 32 and a learning rate of 2e-4. The Adam optimizer is used for network optimization in both two stages. The number of transformer blocks $T$ is set as 4. The hyper-parameter $\gamma$ in Eq. (4) is set to 1. The weighting coefficients $\lambda_1$ and $\lambda_2$ in Eq. (5) are set to 1 and 5, respectively.

**Evaluation Metrics.** To quantify the effectiveness of our method, we utilized three commonly adopted evaluation metrics, including the peak signal-to-noise (PSNR), structural similarity index (SSIM), and normalized mean squared error (NMSE). PSNR values indicate the fidelity and noise levels. SSIM evaluates image similarity based on structure and contrast, considering local and global features. NMSE quantifies average differences between denoised and ground truth images, providing overall accuracy insights. Therefore, higher PSNR and SSIM with lower NMSE suggest better image quality. Note that, we restack 2D slices into entire 3D images for metrics calculation.

### B. Experimental Results

**Comparison with SOTA Methods.** To demonstrate the superiority of the proposed method in leveraging unlabeled and labeled data, we compare our method with five methods including: Cycle-GAN [29], Auto-Context [4], AR-GAN [9], CVT-GAN [11], and SemiPET [22], following the training pipeline in [20]. Table I present the quantitative comparison results. As observed, our method achieves the best performance in terms of all evaluation metrics, demonstrating its superiority. Specifically, compared to the current leading semi-supervised PET reconstruction approach SemiPET, our method still boosts the performance by 0.44 dB PSNR, 0.003 SSIM, and 0.002 NMSE, respectively. In addition, we display the visual comparison results of different methods in Fig. 3. It can be observed that our approach yields the best visual effect, enabling the reconstruction with clearer boundaries and less noise.

TABLE I. QUANTITATIVE COMPARISON RESULTS ON THE UDPET DATASET IN TERMS OF PSNR, SSIM, AND NMSE, AS WELL AS THE PARAMETERS. THE RESULTS ARE PRESENTED AS "MEAN ± STD". THE BEST RESULTS ARE MARKED IN **BOLD**.

| Methods | PSNR [dB]↑ | SSIM↑ | NMSE↓ |
|---|---|---|---|
| LPET | 21.305±1.995 | 0.932±0.024 | 0.080±0.033 |
| Cycle-GAN [27] | 24.203±1.111 | 0.944±0.018 | 0.038±0.011 |
| Auto-Context [4] | 23.067±2.249 | 0.908±0.023 | 0.055±0.030 |
| AR-GAN [9] | 24.411±1.584 | 0.947±0.020 | 0.038±0.013 |
| CVT-GAN [11] | 24.713±1.316 | 0.941±0.020 | 0.035±0.010 |
| SemiPET [20] | 24.755±1.424 | 0.952±0.019 | 0.034±0.011 |
| **S3PET (Ours)** | **25.159±1.570** | **0.955±0.017** | **0.032±0.011** |

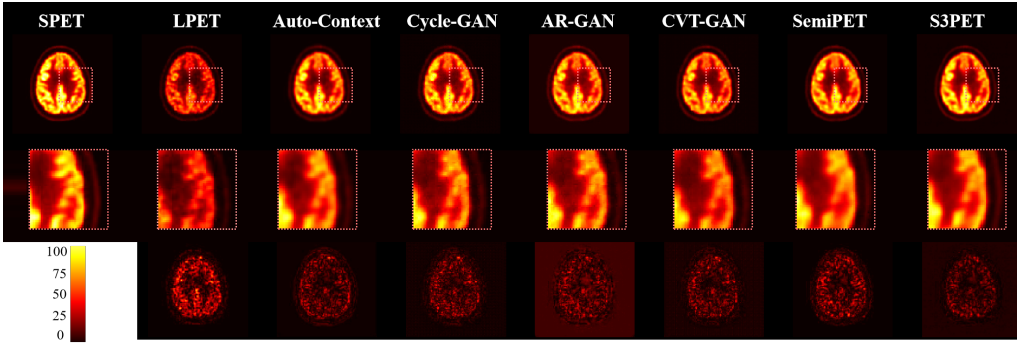

Fig. 3. Visual comparison of the reconstruction results on UDPET dataset. The first row illustrates the SPET, LPET image, and the RPET images obtained from different reconstruction approaches, the second row presents the zoom-in areas of the images for closer inspection, and the third row is the error maps between the reconstructed and the SPET images.

**Ablation Study.** To evaluate the contributions of the key components in the proposed method, we perform the ablation study in a progressive way: (1) leveraging only Stage II with the removal of DKD module and DKL module as the baseline (denoted as Baseline), (2) introducing Stage I to the baseline (denoted as Baseline + DsMAE), (3) injecting DKD to the model in (2) (denoted as Baseline + DsMAE + DKD), (4) employing DKL to the model in (3) (denoted as Baseline + DsMAE + DKD + DKL). The quantitative results are given in Table II. It can be found that the performance boosts progressively with the incorporation of DsMAE, DKD and DKL. Moreover, a significant degradation in the performance can be observed when removing the DsMAE due to the insufficient utilization of abundant unpaired PET images.

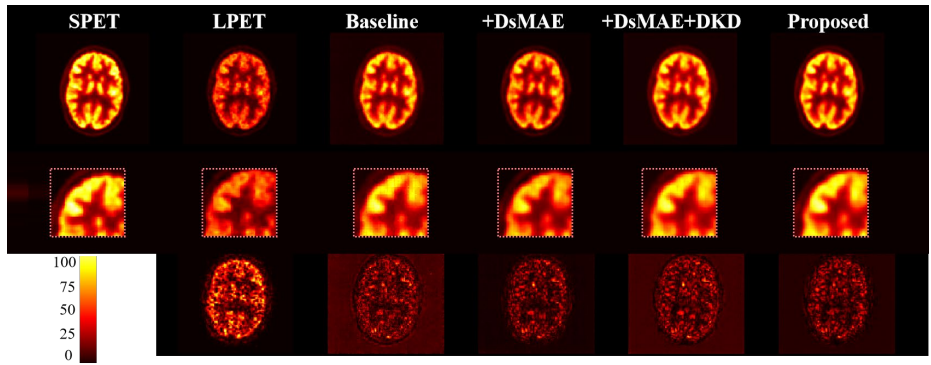

Fig. 4. Visual comparison of the ablation study on UDPET dataset. The first row illustrates the SPET, LPET image, and the RPET images obtained from different ablation variants, the second row presents the zoom-in areas of the images for closer inspection, and the third row is the error maps between the reconstructed and the SPET images.

TABLE II. QUANTITATIVE RESULTS OF THE ABLATION STUDIES ON THE UDPET DATASET IN TERMS OF PSNR, SSIM, AND NMSE. THE RESULTS ARE PRESENTED AS "MEAN ± STD". THE BEST RESULTS ARE MARKED IN **BOLD**.

| Baseline | DsMAE | DKD | DKL | PSNR [dB]↑ | SSIM↑ | NMSE↓ |
|---|---|---|---|---|---|---|
| ✓ | | | | 23.949±1.364 | 0.924±0.016 | 0.041±0.012 |
| ✓ | ✓ | | | 24.659±1.469 | 0.950±0.019 | 0.036±0.012 |
| ✓ | ✓ | ✓ | | 24.763±1.703 | 0.950±0.018 | 0.036±0.014 |
| ✓ | ✓ | ✓ | ✓ | **25.159±1.570** | **0.955±0.017** | **0.032±0.011** |

After adding DsMAE, DKD, and DKL, the results show different degrees of improvement compared to the baseline. Consistent with the quantitative results, the qualitative results presented in Fig. 4 also illustrate that our proposed model achieves the best visual quality. Meanwhile, the removal of components results in more or less noise and artifacts in the reconstructed images. The above results qualitatively and quantitatively suggest the effectiveness of each proposed component in improving the reconstruction quality.

## V. CONCLUSION

In this paper, to tackle the lack of paired training data while avoiding distortions, we innovatively propose a semi-supervised two-stage framework, named S3PET, for high-quality PET image reconstruction. Our model contains an unsupervised pre-training stage and a supervised dose-aware reconstruction stage to allow for an effective utilization of both unpaired and paired PET images during training, thus being able to reconstruct PET images with accurate content and details. Extensive experiments demonstrate the feasibility and superiority of our method.

## REFERENCES

[1] K.A. Johnson, A. Schultz, R.A. Betensky, J.A. Becker, J. Sepulcre, D. Rentz, E. Mormino, J. Chhatwal, R. Amariglio, K. Papp, and G. Marshall, "Tau positron emission tomographic imaging in aging and early Alzheimer disease," *Ann. Neurol.*, vol. 79, no. 1, pp.110-119.

[2] W. Chen, "Clinical applications of PET in brain tumors," J. Nucl. Med., vol. 48, no. 9, pp. 1468-1481, 2022.

[3] Y. Wang, G. Ma, L. An, F. Shi, P. Zhang, D.S. Lalush, X. Wu, Y. Pu, J. Zhou, and D. Shen, "Semisupervised tripled dictionary learning for standard-dose PET image prediction using low-dose PET and multimodal MRI," *IEEE Trans. Biomed. Eng.*, vol. 64, no. 3, pp.569-579, 2016.

[4] L. Xiang, Y. Qiao, D. Nie, L. An, W. Lin, Q. Wang, and D. Shen, "Deep auto-context convolutional neural networks for standard-dose PET image estimation from low-dose PET/MRI," *Neurocomputing*, vol. 267, pp. 406–416, 2017.

[5] Y. Wang, B. Yu, L. Wang, C. Zu, D.S. Lalush, W. Lin, X. Wu, J. Zhou, D. Shen, and L. Zhou, "3D conditional generative adversarial networks for high-quality PET image estimation at low dose," *Neuroimage*, vol. 174, pp.550-562, 2018.

[6] Y. Wang, L. Zhou, B. Yu, L. Wang, C. Zu, D.S. Lalush, W. Lin, X. Wu, J. Zhou, and D. Shen, "3D auto-context-based locality adaptive multi-modality GANs for PET synthesis," *IEEE Trans. Med. Imaging*, vol. 38, no. 6, pp.1328-1339, 2018.

[7] Q. Feng, and H. Liu, "Rethinking PET image reconstruction: Ultra-low-dose, sinogram and deep learning," in *Proc. Int. Conf. Med. Image Comput. Comput.-Assist. Intervent. (MICCAI)*, 2020, pp. 783-792.

[8] K. Spuhler, M. Serrano‑Sosa, R. Cattell, C. DeLorenzo, and C. Huang, "Full-count PET recovery from low-count image using a dilated convolutional neural network," Med. Phys. vol. 47, no. 10, pp. 4928–4938, 2020.

[9] Y. Luo, L. Zhou, B. Zhan, Y. Fei, J. Zhou, Y. Wang,. and D. Shen, "Adaptive rectification based adversarial network with spectrum constraint for high-quality PET image synthesis," *Med. Image Anal.* Vol. 77, pp. 102335, 2022.

[10] Y. Luo, Y. Wang, C. Zu, B. Zhan, X. Wu, J. Zhou, D. Shen, and L. Zhou, "3D transformer-GAN for high-quality PET reconstruction," in *Int. Conf. Med. Image Comput. Comput.-Assist. Intervent. (MICCAI)*, 2021, pp. 276-285.

[11] P. Zeng, L. Zhou, C. Zu, X. Zeng, Z. Jiao, X. Wu, J. Zhou, D. Shen, and Y. Wang, "3D CVT-GAN: A 3d convolutional vision transformer-gan for pet reconstruction," in *Int. Conf. Med. Image Comput. Comput.-Assist. Intervent. (MICCAI)*, Sept. 2022, pp. 516-526.

[12] R. Hu, and H. Liu, "TransEM: Residual swin-transformer based regularized PET image reconstruction," in *Proc. Int. Conf. Med. Image Comput. Comput.-Assist. Intervent. (MICCAI)*, 2022, pp. 184-193.

[13] Y. Zhou, Z. Yang, H. Zhang, I. Eric, C. Chang, Y. Fan, and Y. Xu, "3D segmentation guided style-based generative adversarial networks for pet synthesis," *IEEE Trans. Med. Imaging*, vol. 41, no. 8, pp. 2092-2104, 2022.

[14] J. Cui, P. Zeng, X. Zeng, P. Wang, X. Wu, J. Zhou, Y. Wang, and D. Shen, "TriDo-Former: A Triple-Domain Transformer for Direct PET Reconstruction from Low-Dose Sinograms," in *Proc. Int. Conf. Med. Image Comput. Comput.-Assist. Intervent. (MICCAI)*, pp. 184-194, 2023.

[15] J. Cui, Y. Wang, L. Wen, P. Zeng, X. Wu, J. Zhou, and D. Shen, "Image2Points: A 3D Point-based Context Clusters GAN for High-Quality PET Image Reconstruction," *arXiv preprint* arXiv:2402.00376, 2024.

[16] J. Zhang, Z. Cui, C. Jiang, S. Guo, F. Gao, and D. Shen, "Hierarchical Organ-Aware Total-Body Standard-Dose PET Reconstruction From Low-Dose PET and CT Images," *IEEE Trans. Neural Netw. Learn. Syst.*, 2023.

[17] Z. Han, Y. Wang, L. Zhou, P. Wang, B. Yan, J. Zhou, Y. Wang, and D. Shen, "Contrastive diffusion model with auxiliary guidance for coarse-to-fine PET reconstruction," in *Int. Conf. Med. Image Comput. Comput.-Assist. Intervent. (MICCAI)*, 2023, pp. 239-249.

[18] C. Shen, Z. Yang, and Y. Zhang, "PET Image Denoising with Score-Based Diffusion Probabilistic Models," in *Int. Conf. Med. Image Comput. Comput.-Assist. Intervent. (MICCAI)*, 2023, pp. 270-278.

[19] C. Jiang, Y. Pan, M. Liu, L. Ma, X. Zhang, J. Liu, X. Xiong, and D. Shen, "PET-Diffusion: Unsupervised PET Enhancement Based on the Latent Diffusion Model," in *Int. Conf. Med. Image Comput. Comput.-Assist. Intervent. (MICCAI)*. pp. 3-12, 2023.

[20] C. Niu, W. Cong, F.L. Fan, H. Shan, M. Li, J. Liang, and G. Wang, "Low-Dimensional Manifold-Constrained Disentanglement Network for Metal Artifact Reduction," *IEEE Trans. Radiat. Plasma Med. Sci.* vol. 6, no. 6, pp. 656-666, 2021.

[21] T. Wang, H. Yu, Z. Wang, H. Chen, Y. Liu, J. Lu, and Y. Zhang, "SemiMAR: Semi-Supervised Learning for CT Metal Artifact Reduction," *IEEE J. Biomed. Health. Inform.*, 2023.

[22] C. Jiang, Y. Pan, Z. Cui, D. Nie, and D. Shen, "Semi-supervised Standard-dose PET Image Generation via Region-adaptive Normalization and Structural Consistency Constraint," *IEEE Trans. Med. Imaging*, vol. 42, pp. 2974-2987, 2023.

[23] Y. Fei, C. Zu, Z. Jiao, X. Wu, J. Zhou, D. Shen, and Y. Wang, "Classification-Aided High-Quality PET Image Synthesis via Bidirectional Contrastive GAN with Shared Information Maximization," in *Proc. Int. Conf. Med. Image Comput. Comput.-Assist. Intervent. (MICCAI)*, pp. 527-537, 2022.

[24] K. He, X. Chen, S. Xie, Y. Li, P. Dollár, and R. Girshick, "Masked autoencoders are scalable vision learners," in *Proceedings of the IEEE/CVF conference on computer vision and pattern recognition (CVPR)*, pp. 16000-16009, 2022.

[25] L. Zhou, H. Liu, J. Bae, J. He, D. Samaras, and P. Prasanna, "Self pre-training with masked autoencoders for medical image classification and segmentation," in *International Symposium on Biomedical Imaging (ISBI)*, 2023, pp. 1-6.

[26] T. Yoon, and D. Kang, "Enhancing pediatric pneumonia diagnosis through masked autoencoders," *Sci. Rep.*, vol. 14, no. 1, pp.6150, 2023.

[27] D. Wang, Y. Xu, S. Han, and H. Yu, "Masked Autoencoders for Low-dose CT Denoising," in *International Symposium on Biomedical Imaging (ISBI)*, pp. 1-4.19, 2023.

[28] MICCAI challenges: Ultra-low dose pet imaging challenge 2022. https://doi.org/10.5281/zenodo.6361846 (2022)

[29] J.Y. Zhu, T. Park, P. Isola, and A.A. Efros, "Unpaired image-to-image translation using cycle-consistent adversarial networks," in *Proceedings of the IEEE/CVF conference on computer vision and pattern recognition (CVPR)*, pp. 2223-2232, 2017.